\documentclass[11pt,twoside]{article}
\usepackage[dvips]{graphicx}
\usepackage{natbib}
\usepackage{fancyhdr,setspace,subfigure,rotate}
\usepackage{amsthm,amssymb,amsmath}
\usepackage{lineno}
\usepackage{color}
\usepackage{ctable}
\IfFileExists{url.sty}{\usepackage{url}}{\newcommand{\url}{\text}}
\usepackage{sectsty}
\allsectionsfont{\normalsize}
\numberwithin{equation}{section}

\begin{document}
\title{\large\bfseries A BAYESIAN ESTIMATION FOR THE FRACTIONAL ORDER OF THE DIFFERENTIAL EQUATION THAT MODELS TRANSPORT IN UNCONVENTIONAL HYDROCARBON RESERVOIRS }
\author{
J.~Whitlinger$\!^a$, E.L.~Boone$\!^a$, and R.~Ghanam$^b$ \\[0.2in] 
\normalsize\itshape $^a$Department of Statistical Sciences and Operations Research,\\ 
\normalsize\itshape Virginia Commonwealth University, Richmond, VA 23284, USA\\[0.05in]
\normalsize\itshape $^b$Virginia Commonwealth University - Qatar,\\ 
\normalsize\itshape Education City, Doha, Qatar.\\
}

\date{\today}
\maketitle

\hrule
\begin{abstract}
The extraction of natural gas from the earth has been shown to be governed by differential equations concerning flow through a porous material. Recently, models such as fractional differential equations have been developed to model this phenomenon. One key issue with these models is estimating the fraction of the differential equation. Traditional methods such as maximum likelihood, least squares, and even method of moments are not available to estimate this parameter as traditional calculus methods do not apply. We develop a Bayesian approach to estimate the fraction of the order of the differential equation that models transport in unconventional hydrocarbon reservoirs. In this paper, we use this approach to adequately quantify the uncertainties associated with the error and predictions. A simulation study is presented as well to assess the utility of the modeling approach.
\end{abstract}
\vspace{0.1in}
\hrule

\section{Introduction}
\label{sect-intro}

Fractional calculus and its application to different disciplines of science has grown. Recently, varied works have been published about fractional differential equations for a range of topics, from the case of L\'evy motion discussed by \cite{Benson}, to the modeling of an Ebola epidemic in \cite{area}. Fractional calculus has also been discussed to great length in \cite{meer}.

\cite{wyss} discussed fractional derivative concepts, including various ways to solve them. \cite{LiuZhan} considered a finite domain space-time fractional advection dispersion equation. \cite{MalGh} discuss the lack of knowledge of transport in porus material. The use of fractional differential equations is becoming more popular in the field of hydrocarbon reservoirs, as it utilizes all the parameters previously stated. \cite{deA} discuss anomalous flow necessitating the use for fractional calculus, specificaly fractional diffusion equation. In this paper, the goal is to estimate the fractional order $\alpha$ of equation (\ref{sect-intro}.\ref{eq:malik}) which is difficult to isolate. \cite{fan} discuss parameter estimation of a fractional fractal diffusion model, specifically across porous media. \cite{Malik} considered the problem of modeling transport in unconventional hydrocarbon reservoirs using a fractional partial differential equation. \cite{awotunde} show there is deviation between real data obtained and the model using Darcy$'$s law from \cite{darcy}, which is why new models must be implemented. \cite{ariza} utilize a Bayesian approach to solve the inverse problem of a fractional population growth model. They employ the \cite{plum} JAGS (Just Another Gibbs Sampler) software to generate Markov Chain Monte Carlo (MCMC) samples, which for more information on MCMC, see \cite{gilks}. While JAGS is effective, it cannot reliably be guaranteed to produce quality samples. This work utilizes a Sampling Importance Resampling approach to obtain the posterior samples giving the user more control over the quality of the sample as well as ensuring the differential equation model is solved correctly. Furthermore, this work is applied to hydrocarbon reservoirs. Other publications, such as \cite{raz}, discuss fractional calculus use in fractured reservoirs.

The time-fractional advection-diffusion equation is given by \cite{Malik}:
\begin{equation}
\frac{ \partial^{ \alpha } p}{ \partial t^{\alpha} } = \frac{ \partial}{ \partial x} \left( K \frac{ \partial p }{\partial x} \right) - U  \left( \frac{ \partial p }{\partial x} \right), ~~~~~~~~t>0, 0 \le x \le 1
\end{equation}\label{eq:malik}

\noindent where $p(x,t)$ is the pressure distribution in unconventional reservoirs,  $K=K(p)$ is the diffusivity (which is related to the rock permeability), and  $U=U(p,p_x)$ is a convection velocity; both $K$ and $U$ are highly non-linear. 

Observed data rarely follows the solution exactly, as realizations are often contaminated by some source of error. The question becomes: where are the sources of error? There are two sources of error, internal error (also known as noise or process error) and external error (also known as measurement or experimental error). \cite{Strong} state that internal error is minor misspecification of the model across the process space, specifically in the homogeneity of the substrate. With external error, measuring instruments often fail to yield the same measurement twice. 

Internal error:

\[ \frac{ \partial^{ \alpha } p}{ \partial t^{\alpha} } = \frac{ \partial}{ \partial x} \left( K \frac{ \partial p }{\partial x} \right) - U  \left( \frac{ \partial p }{\partial x} \right) + \mathbf{ \epsilon } \]

\noindent Here, internal error means the error in the differential equation model.  In this case, the internal error will accumulate over the integral associated with the solution.  Hence, $\epsilon$ may be something like a Brownian motion or  Ornstein-Uhlenbeck process.\\

External error:

\[ p_i^*(x,t) = p(x,t) +  \epsilon_i  \]

\noindent where $ p_i^*(x,t) $ is the $i^{th}$ observed value of the process $p(x,t)$ at time $t$ and $x$.  In this case, $\epsilon_i$ will follow some appropriate probability distribution.  The external error problem is an easy problem in contrast to the internal error problem. For the proof of concept, the model will only focus on external error. Ultimately, both types of error structures in the model would be ideal.

\section{Model}
The fractional differential equation of interest is taken from \cite{Malik} to model the pressure as a function of time $t$ and location $x$ under exponential uploading, given by $p(x,0)=e^{-cx}$, ($c > 0$). In this case, $p(x,t)$ has a closed form solution under some mild conditions. 
\begin{equation}
p(x,t) = e^{-x} \sum_{k=0}^{\infty} \frac{2t^{\alpha k} }{\Gamma (\alpha k + 1) }
\end{equation}\label{eq:pxtnoerror}

\noindent The main parameter in this model is $\alpha$, the fraction of the derivative, and it is bound between 0 and 1.

\subsection{Statistical Model and Parameter Estimation}
A Bayesian approach is used to estimate the parameters in the model which requires that both a likelihood be specified as well as prior distributions on the model parameters:

\begin{equation}
p_i(x,t) = \left[ e^{-x} \sum_{k=0}^{\infty} \frac{2t^{\alpha k} }{\Gamma (\alpha k + 1) } \right] \epsilon_{i}(x,t)
\end{equation}\label{eq:likelihood}

where $\epsilon_{i}(x,t) \overset{iid}{\sim} LogNormal(1,\sigma^2)$. Here, the logNormal likelihood is chosen to ensure that pressure is always a positive value. In this specification, the likelihood has two parameters, $\alpha$ and $\sigma^2$. The prior distributions are specified as $\alpha \sim Beta \left( \alpha^*,\beta^* \right)$ to reflect the prior knowledge that $\alpha$ is bound between 0 and 1. For $\sigma^2$ the prior distribution is specified as $\sigma^2 \sim \chi^2 (df)$ to reflect the prior knowledge that $\sigma^2$ must be a positive value. For more on prior distribution and selection, see \cite{berger}.

The posterior distribution $\pi \left( \alpha, \sigma^2 | x, t, p_i(x,t) \right)$ can be found using Bayes' Theorem:
\begin{equation}
\pi \left( \alpha, \sigma^2 | x, t, p_i(x,t) \right)= \frac{ \pi(\alpha, \sigma^2) L( p_i(x,t) | x, t, \alpha, \sigma^2)}{  \int \pi(\alpha, \sigma^2) L( p_i(x,t) | x, t, \alpha, \sigma^2) d\alpha d\sigma^2 }
\end{equation}
In the case considered here, there is no analytic solution to $\pi \left( \alpha, \sigma^2 | x, t, p_i(x,t) \right)$. Thus, a sampling method must be employed to draw samples from the posterior distribution, from which all inferences will be made.  There are many choices for the algorithm to sample from the posterior distribution such as Acceptance Sampling, Metropolis-Hastings Sampling, Sampling Importance Resampling, etc, found in \cite{gelman} and \cite{gilks}.  Due to the low number of model parameters, the Sampling Importance Resampling approach is employed in this work.  For generality of notation, let $\theta = (\alpha, \sigma^2)$.

\textbf{Algorithm}\\

\begin{enumerate}
  \item Draw $\theta_1, \theta_2, ... , \theta_{n_c} \sim p_c (\theta)$ where $p_c(\theta)$ is a distribution similar to the posterior distribution from which it is easy to draw candidate values.
  \item Calculate $w_i = \frac{ \pi(\theta_i)L( p_i(x,t)  |  \theta_i, x, t) }{ p_c(\theta_i) }$
  \item Calculate the posterior probability weights $w_i^{*} = \frac{ w_i }{ \sum_{i=1}^{n_c} w_i }$.
  \item Resample $n_s$ samples with replacement from $\theta_1,\theta_2,...,\theta_{n_c}$ using their corresponding posterior probability weights $w_i^{*}$.
  \item The set  of posterior samples of size $n_s$ is obtained by taking $n_s$ samples, with replacement, from $\theta_1,\theta_2,...,\theta_{n_c}$ using their corresponding posterior probability weights $w_i^{*}$.
\end{enumerate}

\section{Simulated Example}
\begin{figure}
  \begin{center}
  \begin{tabular}{c c}
    \includegraphics[width=0.5\textwidth]{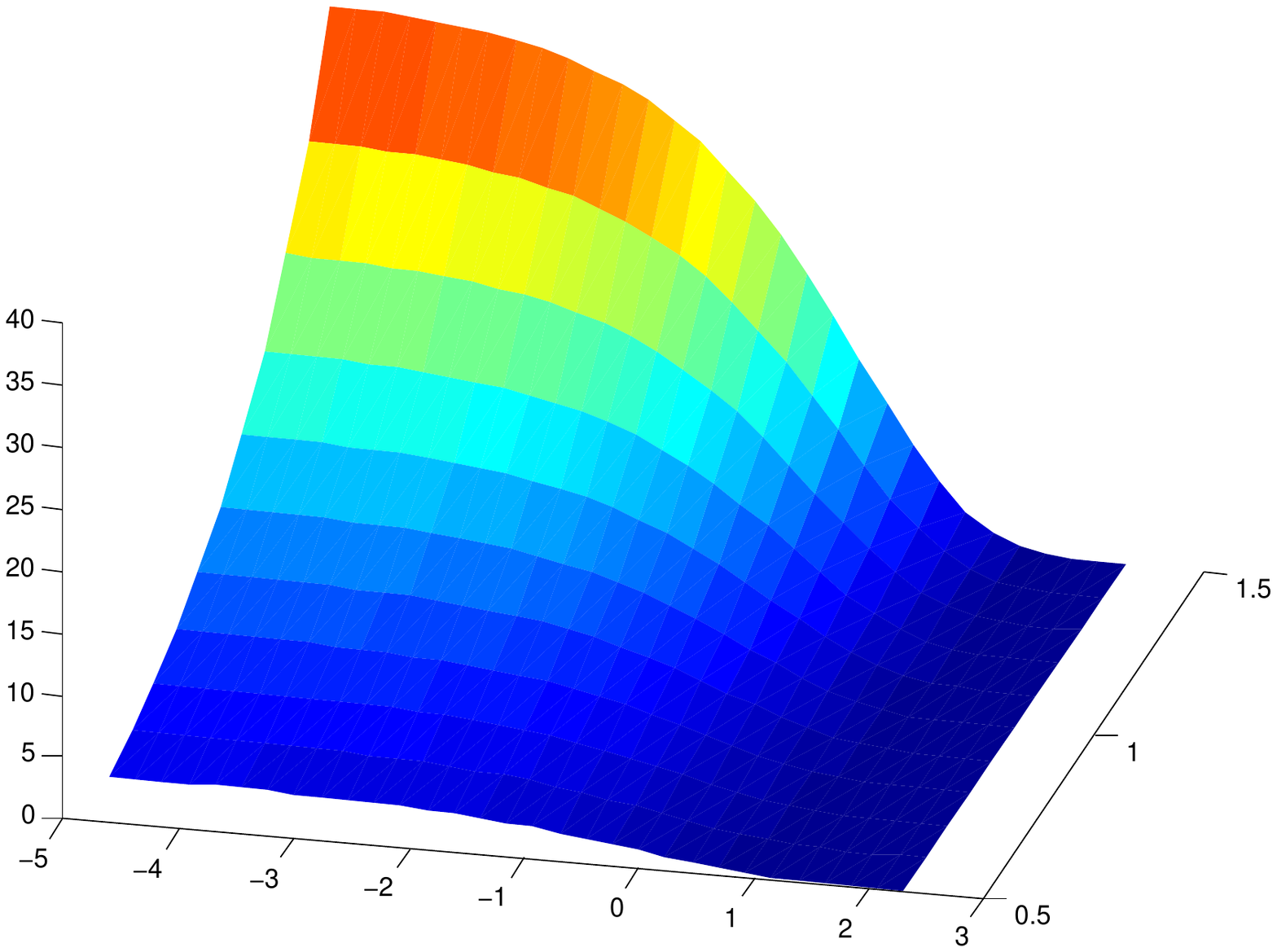} &
    \includegraphics[width=0.5\textwidth]{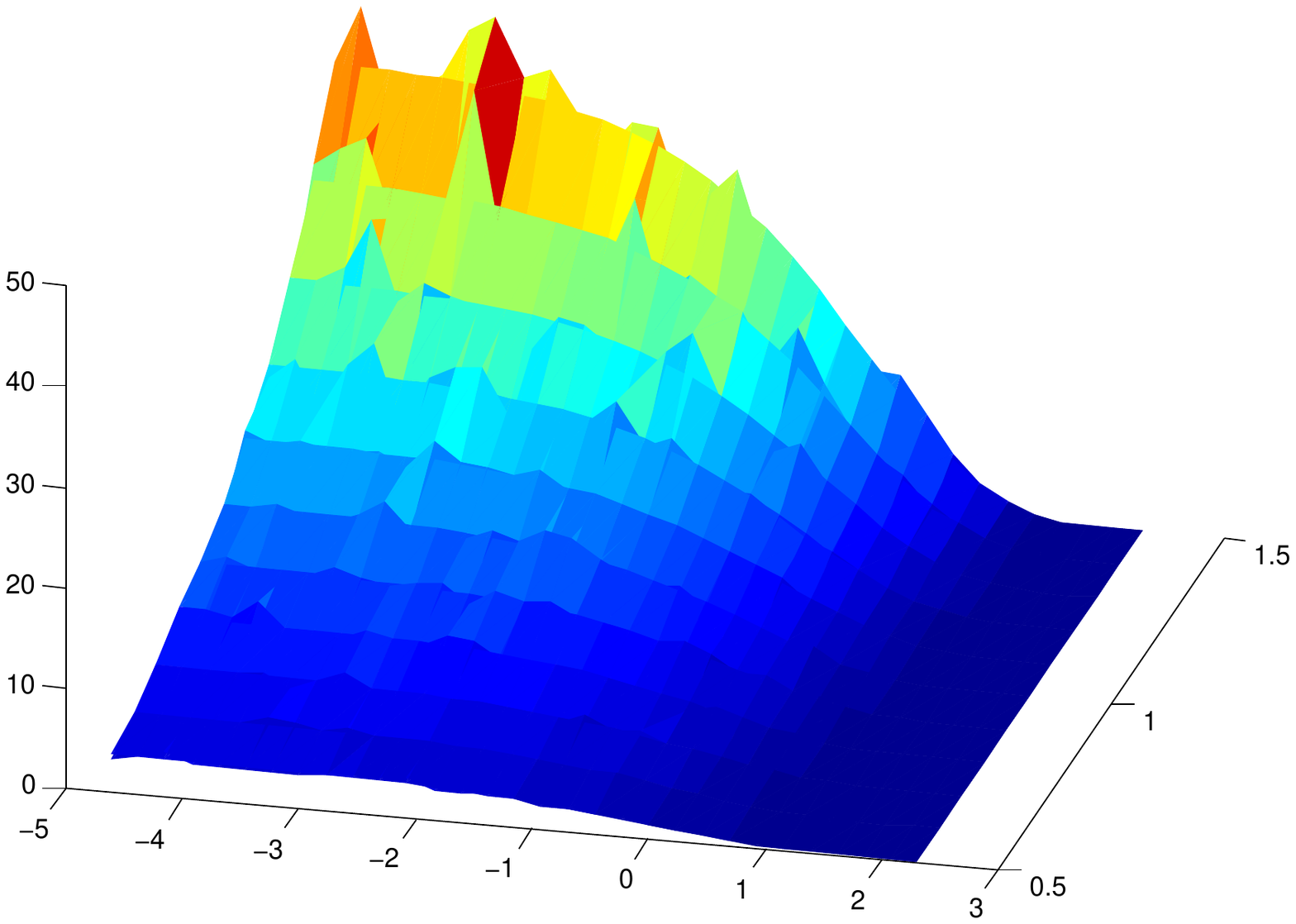}
    \end{tabular}
  \end{center}
  \caption{Pure data (left) from (\ref{sect-intro}.\ref{eq:pxtnoerror}) and Noisy data (right) using the same surface with logNormal noise $\sigma=0.1$.} 
\label{fig:PureNoise1}
\end{figure}

Suppose the system of interest is given by (\ref{sect-intro}.\ref{eq:pxtnoerror}) where $\alpha = 0.82$.  Further, suppose that data for $p(x,t)$ has been observed, with noise, at all combinations of $31$ equally spaced levels of $x$ from $0.01$ to $10$ and $11$ equally spaced times $t$ from $0.5$ to $1.5$ and the noise is multiplicative following a LogNormal distribution with mean 1 and $\sigma = 0.1$.  Figure~\ref{fig:PureNoise1} shows the unperturbed data surface on the left and the perturbed data surface on the right.  Notice that this set parameter specification produces a quite noisy surface.

The prior distributions for $\alpha$ and $\sigma$ were specified as follows:
\begin{eqnarray}
  \alpha    &\sim& Beta( 3, 3 )  \nonumber \\
  \sigma^2 &\sim& \chi^2(1) \nonumber.
\end{eqnarray}
Using the LogNormal likelihood, the prior above, and Bayes Formula, the posterior distribution is given as:

\begin{equation}
\pi \left( \alpha, \sigma^2 | x, t, p_i(x,t), \mu_i(x,t) \right)= \propto \alpha^2(1-\alpha)^2 \prod_{i=1}^n \frac{e^{-\frac{(ln p_i(x,t)-\mu_i(x,t)^2)}{2\sigma^2}}}{p_i(x,t)\sigma\sqrt{2 \pi}}
\end{equation}

\noindent where $p_i(x,t)$ is the realized value and $\mu_i(x,t)$ is the theoretical without noise. For clarity, $\mu_i(x,t)$ is the solution to the system for $x,t$ for the $i^{th}$ sampling point.

\begin{figure}
  \begin{center}
  \begin{tabular}{c c}
    \includegraphics[width=0.5\textwidth]{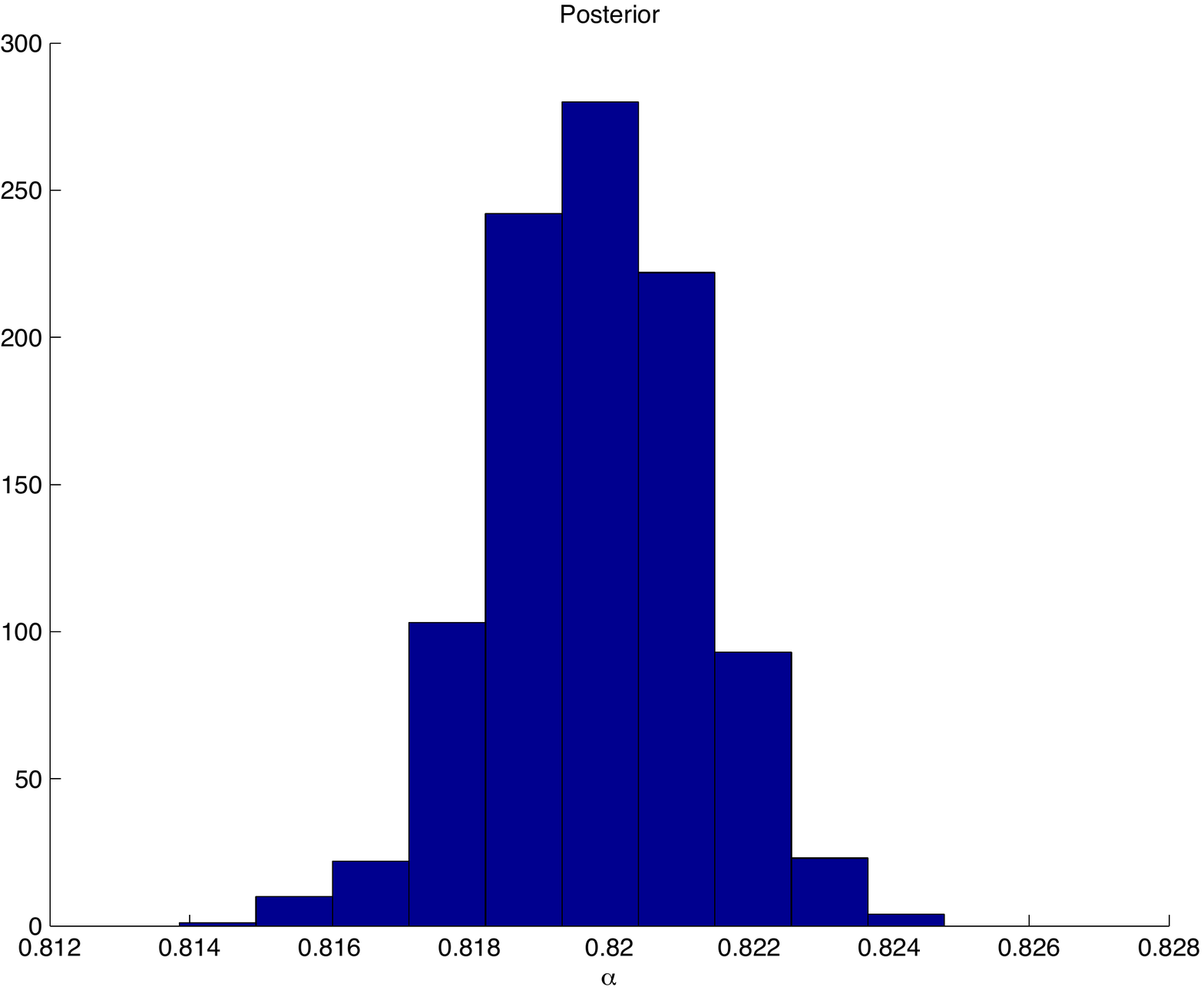} &
    \includegraphics[width=0.5\textwidth]{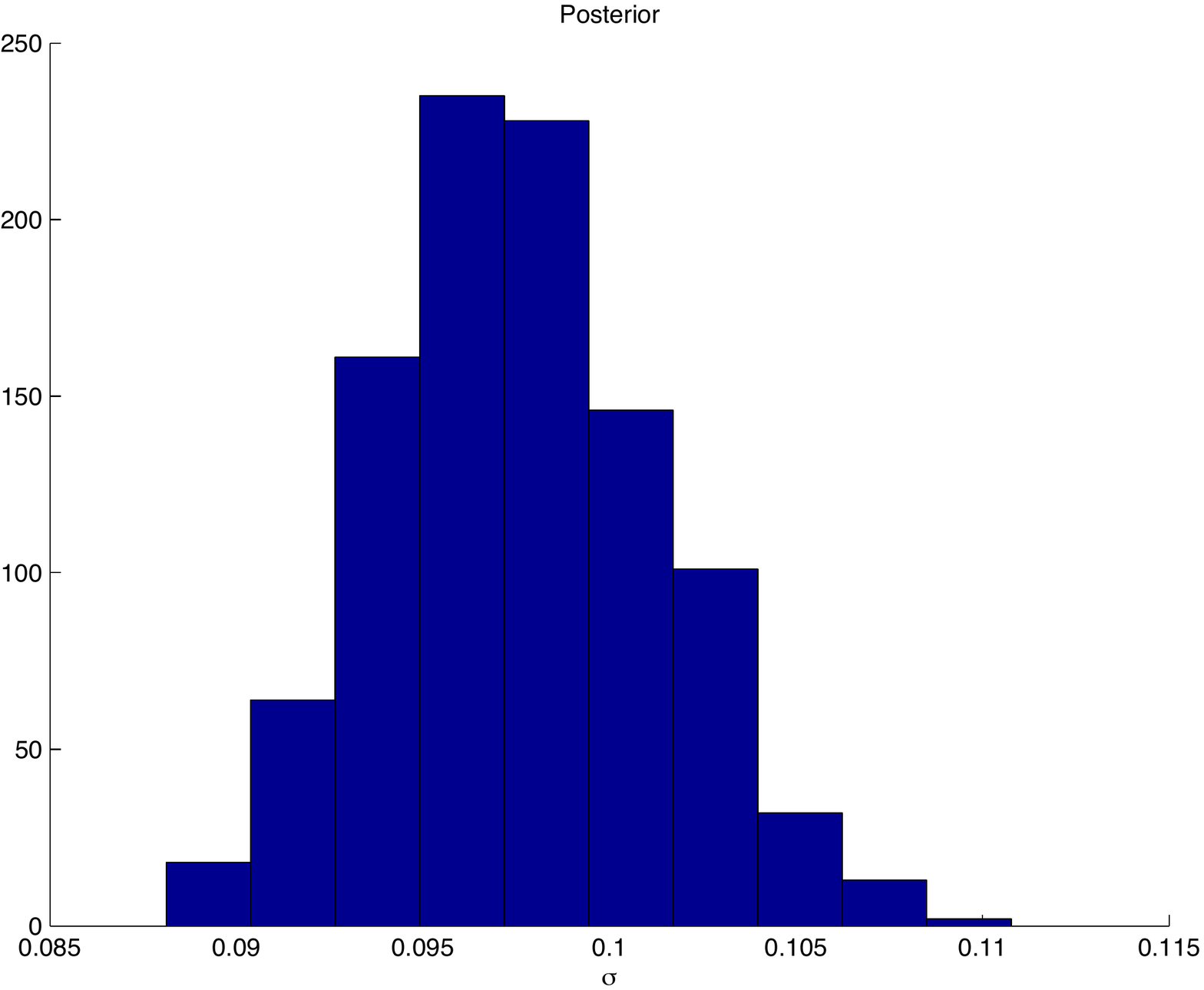}
    \end{tabular}
  \end{center}
  \caption{Histograms of the marginal posterior distributions of $\alpha$ (left) and $\sigma$ (right).} 
\label{fig:AlphSigPost}
\end{figure}

The Sampling Importance Resampling algorithm was employed with 10,000 candidate samples with 1,000 posterior samples drawn.  The sample generated a posterior sample of 89.8\% unique samples indicating a high quality sample from the posterior distribution.  Histograms of the posterior distributions of $\alpha$ and $\sigma$ can be found in Figure~\ref{fig:AlphSigPost}, which shows the true values $\alpha = 0.82$ and $\sigma = 0.1$ in the middle of the distribution.  For reference, the 95\% posterior credible intervals were generated by taking the 2.5\% and 97.5\% quantiles from the marginal posterior distributions and gave: for $\alpha$, (0.8169, 0.8226), which contains the true value 0.82, and for $\sigma$, (0.0907, 0.1050), which also contains the true value of 0.1.  This shows that the procedure proposed can be employed to use data to estimate the fraction of the differential equation.

\begin{figure}
  \begin{center}
  \begin{tabular}{c c c}
    \includegraphics[width=0.33\textwidth]{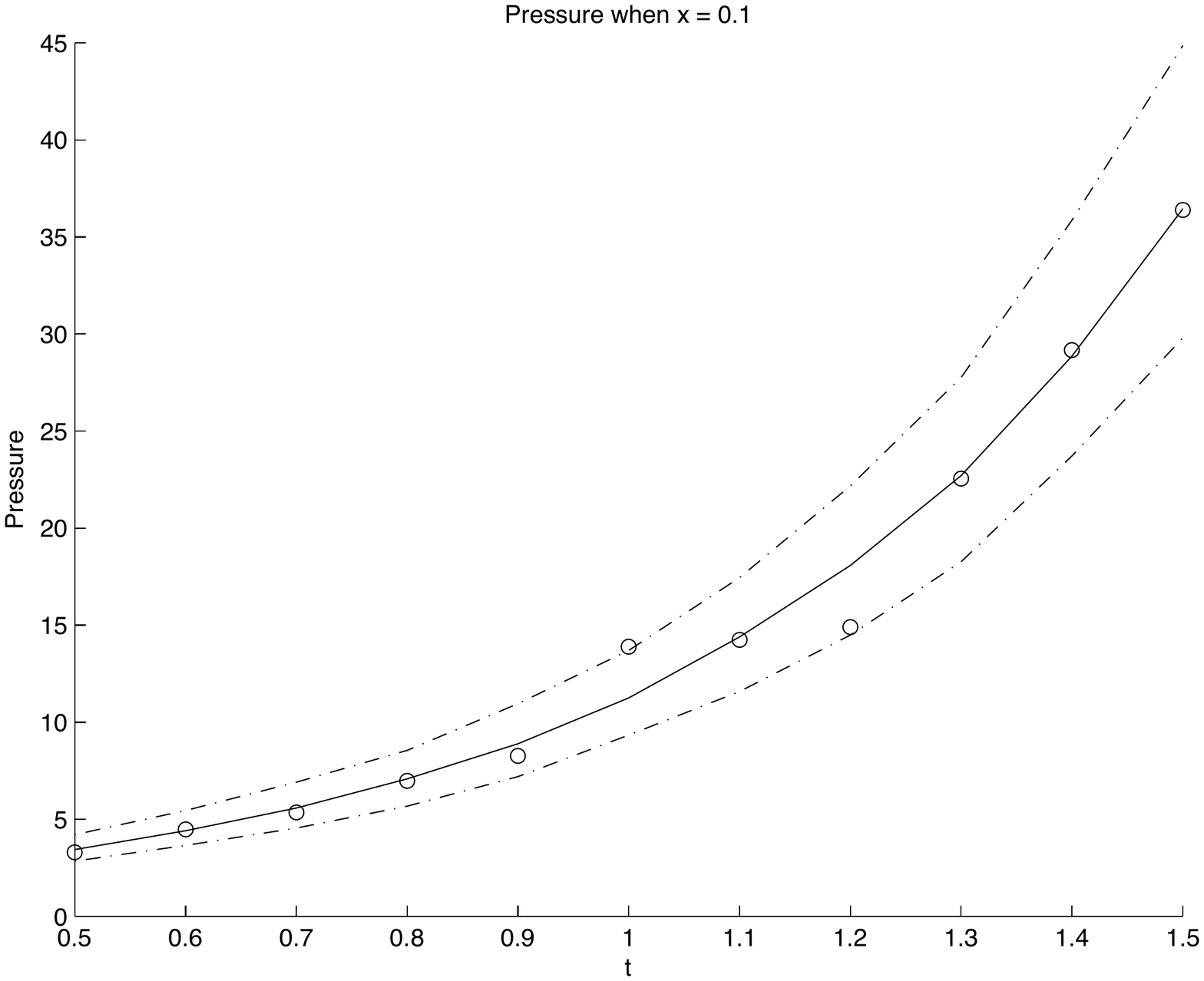} &
    \includegraphics[width=0.33\textwidth]{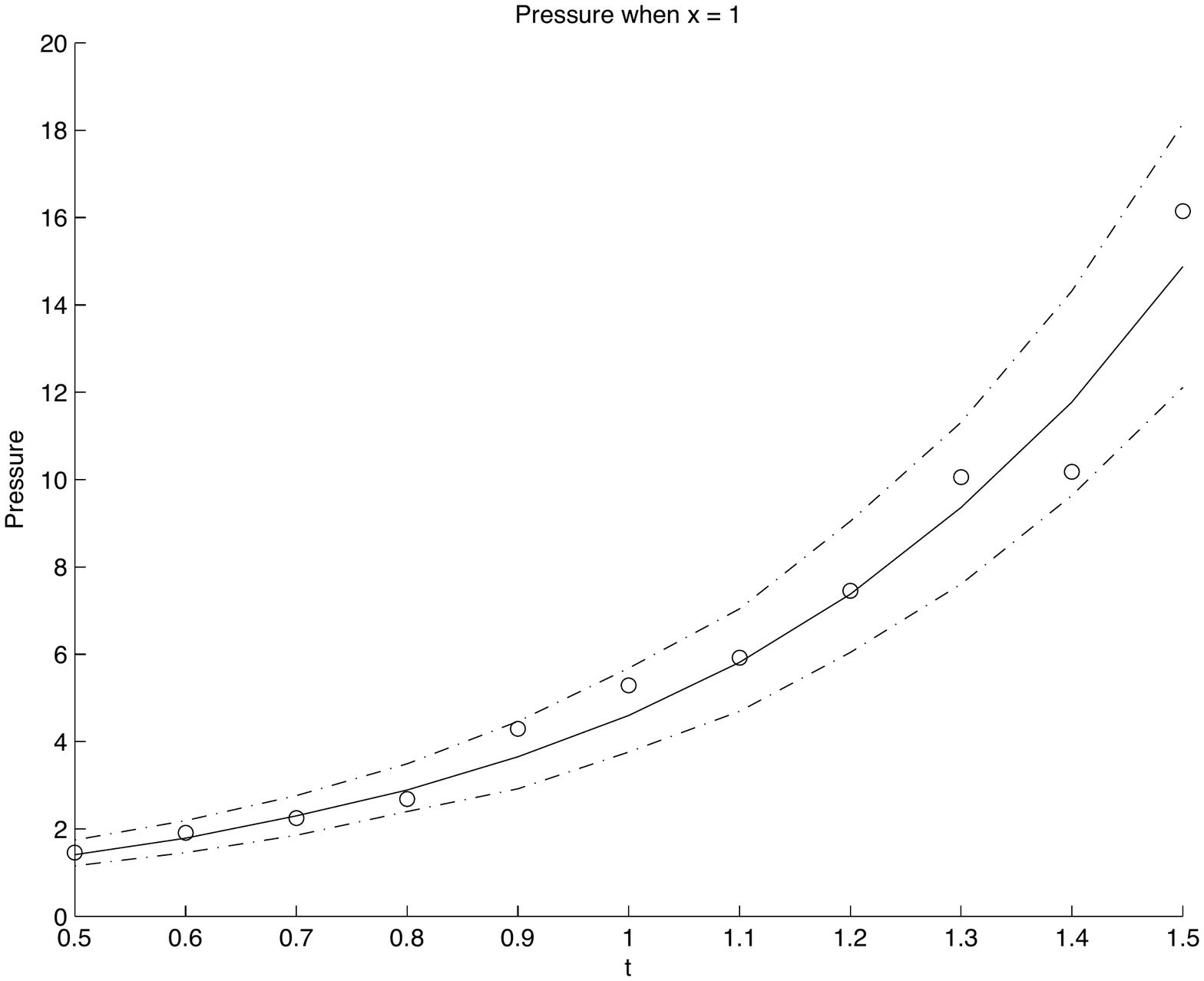} &
    \includegraphics[width=0.33\textwidth]{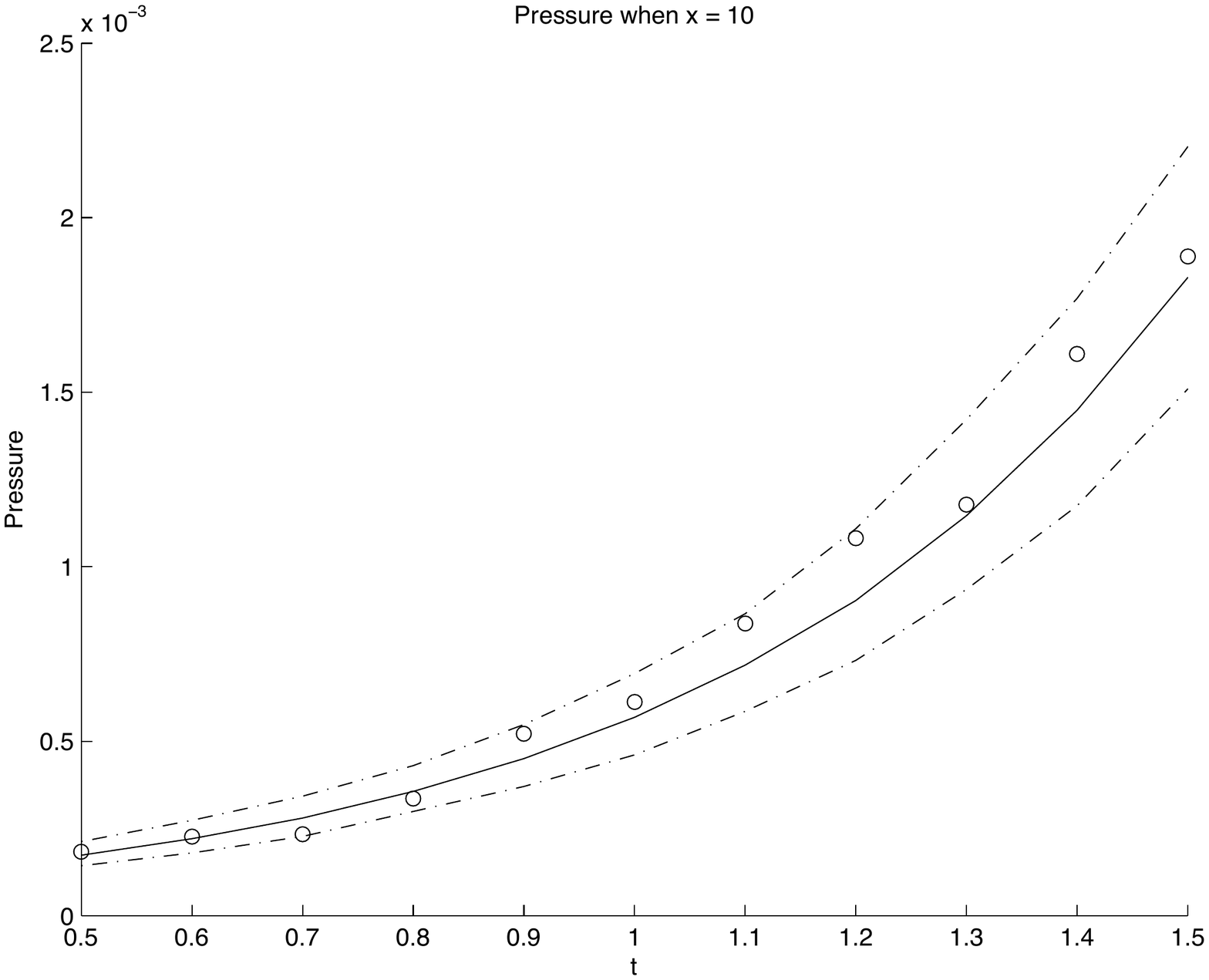}
    \end{tabular}
  \end{center}
  \caption{Profiles of predictive distribution with 95\% predictive intervals across $t$.  Data values ($\circ$), median ($-$), $Q_{2.5}$ and $Q_{97.5}$ ($- \cdot -$). } 
\label{fig:Xslice}
\end{figure}

\begin{figure}
  \begin{center}
  \begin{tabular}{c c c}
    \includegraphics[width=0.33\textwidth]{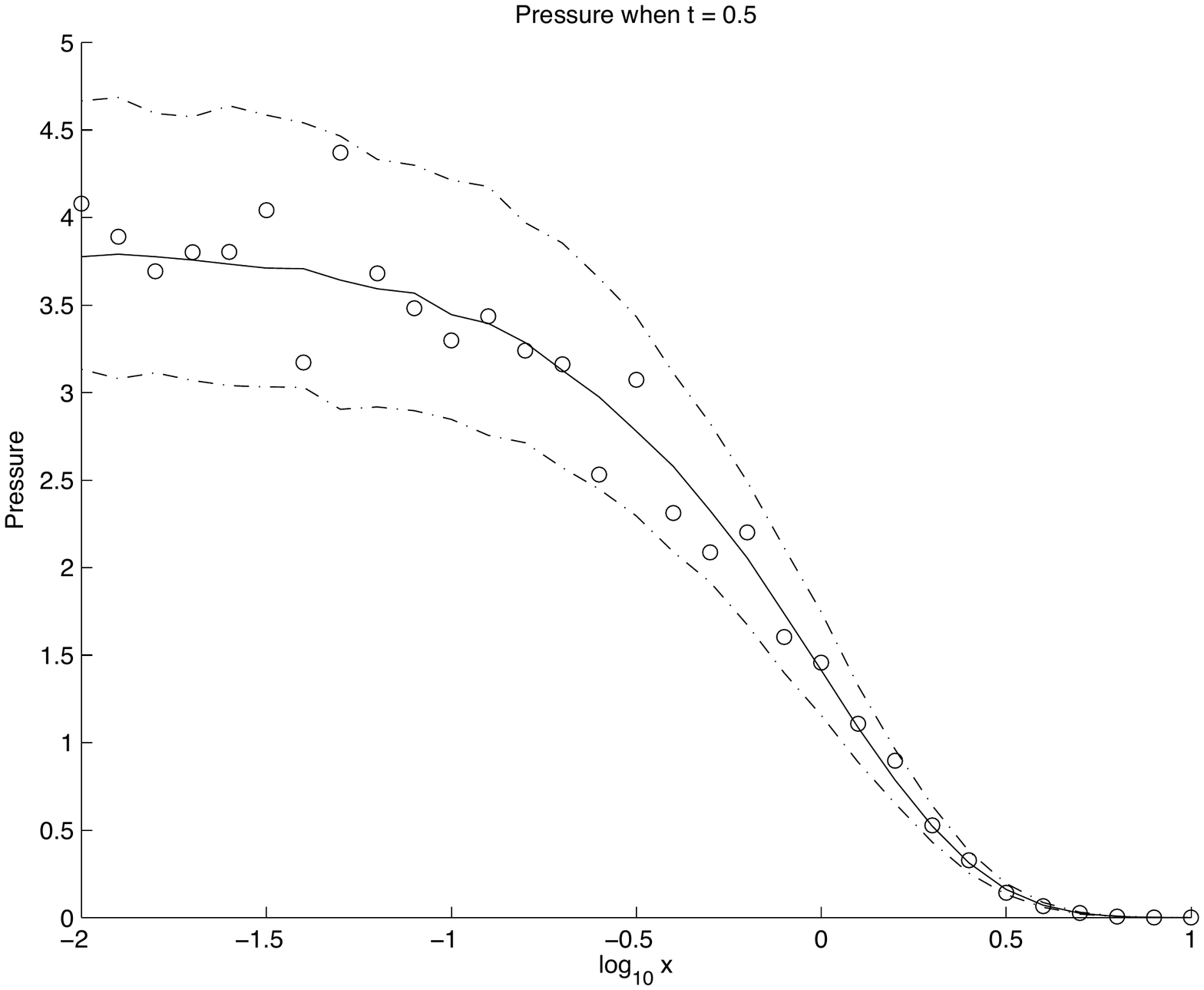} &
    \includegraphics[width=0.33\textwidth]{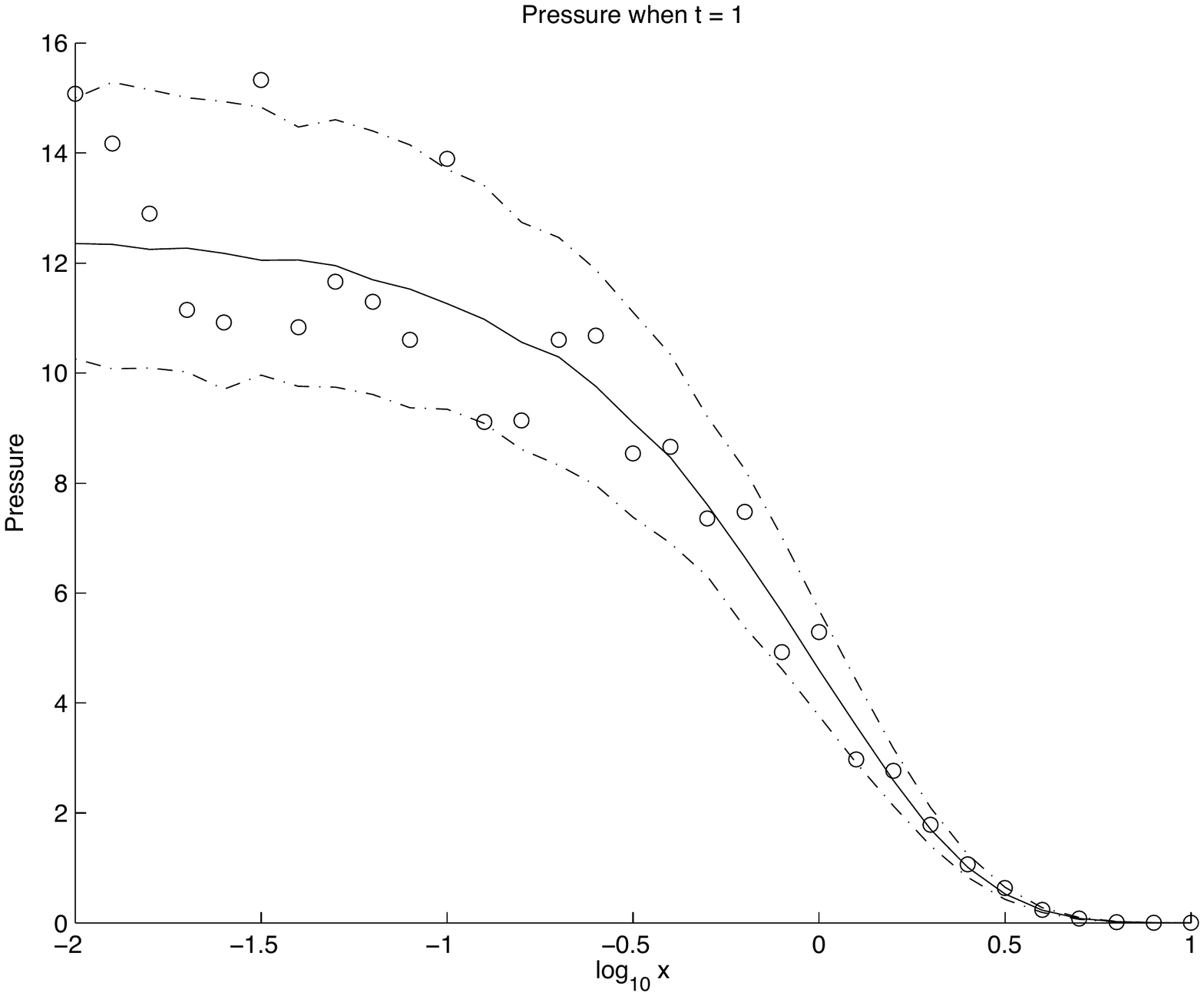} &
    \includegraphics[width=0.33\textwidth]{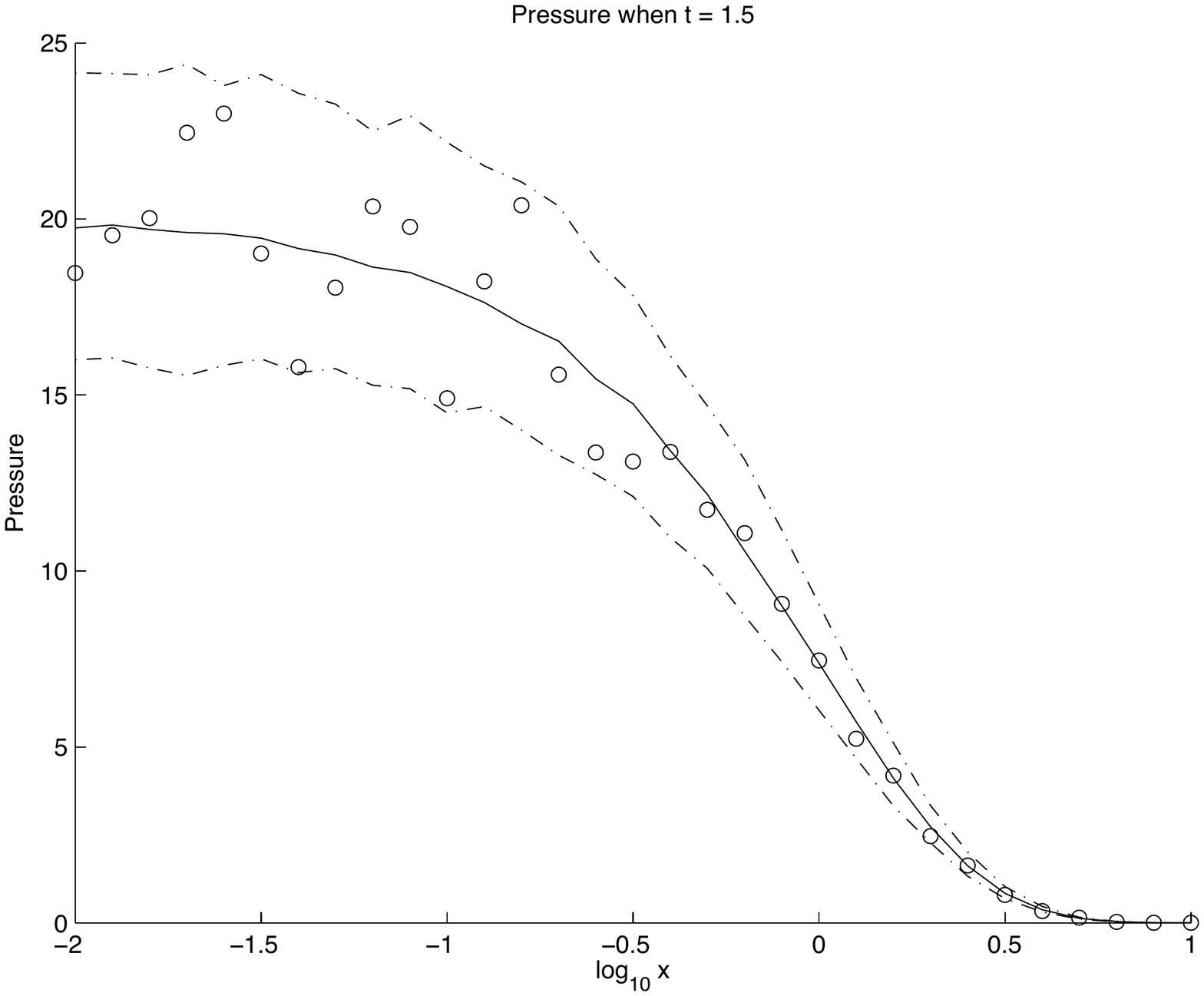}
    \end{tabular}
  \end{center}
  \caption{Profiles of predictive distribution with 95\% predictive intervals across $x$. Data values ($\circ$), median ($-$), $Q_{2.5}$ and $Q_{97.5}$ ($- \cdot -$).} 
\label{fig:Tslice}
\end{figure}

Not only can the method proposed be used to estimate the fraction of the differential equation, it can also be used to quantify the prediction uncertainty associated with the model and parameter estimates.  To do this, the posterior predictive distribution can be employed to generate a distribution for a new observation $p(x_{new},t_{new})$ at the value of $x_{new}$ and $t_{new}$.  Recall that the posterior predictive distribution is given by:
\begin{equation}
\pi( p(x_{new},t_{new} | p_i(x,t), x, t ) = \int \pi(\alpha, \sigma^2 |  p_i(x,t), x, t, \alpha, \sigma^2) L( p(x_{new},t_{new}) |  p_i(x,t), x, t, \alpha, \sigma^2) d\alpha d\sigma^2.
\end{equation}
In the case considered here, the posterior predictive distribution is difficult to visualize in three dimensions.  Instead, profile plots of the median surface are created, the surfaces generated by the 2.5\% and 97.5\% quantiles for given values of $t$ and given values of $x$ along with the data.  Figures~\ref{fig:Xslice} and~\ref{fig:Tslice} give these profile plots for $x$ and $t$, respectively.  Notice that the posterior predictive intervals capture most of the observed data.  This gives evidence that the modeling approach is properly quantifying the uncertainties associated with both estimation as well as inherent noise in the data.

\section{Robustness}
In order to determine if the approach proposed for estimating $\alpha$ is robust to the value of $\alpha$ in the underlying process, a robustness analysis was conducted for varying values of $\alpha$ and $\sigma$.  For this study, a new dataset was simulated using each combination of $\alpha = 0.1$ to $0.9$ by increments of $0.1$ and $\sigma = 0.01, 0.1$ and $0.25$.  With these simulated datasets the estimation algorithm was used to estimate the 95\% credible intervals for each parameter using the 2.5\% and 97.5\% quantiles from the 1,000 samples from the posterior distribution of $\alpha$ and $\sigma$.  Table~\ref{tbl:robust} shows the results of this study.  Notice that all of the 95\% credible intervals contain the correct parameter value, which indicates that the estimation algorithm is robust to the underlying values of $\alpha$ and $\sigma$.

\begin{table}
  \begin{center}
   \caption{Robustness of estimating $\alpha$ and $\sigma$ study for simulated datasets with $\alpha = 0.1$ to $0.9$ in increments of $0.1$ and $\sigma = 0.01, 0.1$ and $0.25$. Quantiles of the posterior distribution were used to create 95\% credible intervals for $\alpha$ and $\sigma$, $(Q_{0.025},Q_{0.975})$. } 
    \begin{tabular}{ c c |  c c c  } \hline
     $\alpha$ &                         &  $\sigma = 0.01$ &  $\sigma = 0.1$  &  $\sigma = 0.25$ \\ \hline
     0.1         & $\hat{\alpha}$  & (0.0999, 0.1001) & (0.0999, 0.1002) & (0.0999, 0.1002) \\
                   & $\hat{\sigma}$ & (0.0090, 0.0104) & (0.0912, 0.1065) & (0.2461, 0.2838)  \\
     0.2         & $\hat{\alpha}$  & (0.1999, 0.2001) & (0.1999, 0.2001) & (0.1998, 0.2001) \\
                   & $\hat{\sigma}$ & (0.0094, 0.0110) & (0.0916, 0.1067) & (0.2356, 0.2725)  \\
     0.3         & $\hat{\alpha}$  & (0.2999, 0.3001) & (0.2998, 0.3002) & (0.2995, 0.3001) \\
                   & $\hat{\sigma}$ & (0.0093, 0.0108) & (0.0886, 0.1022) & (0.2322, 0.2699)  \\
     0.4         & $\hat{\alpha}$  & (0.3999, 0.4001) & (0.3997, 0.4005) & (0.3984, 0.4003) \\
                   & $\hat{\sigma}$ & (0.0098, 0.0114) & (0.0868, 0.1003) & (0.2245, 0.2603)  \\
     0.5         & $\hat{\alpha}$  & (0.4998, 0.5001) & (0.4990, 0.5006) & (0.4980, 0.5016) \\
                   & $\hat{\sigma}$ & (0.0094, 0.0109) & (0.0987, 0.1146) & (0.2175, 0.2544)  \\
     0.6         & $\hat{\alpha}$  & (0.5998, 0.6002) & (0.5986, 0.6015) & (0.5970, 0.6035) \\
                   & $\hat{\sigma}$ & (0.0096, 0.0112) & (0.0987, 0.1161) & (0.2276, 0.2629)  \\
     0.7         & $\hat{\alpha}$  & (0.6999, 0.7003) & (0.6996, 0.7038) & (0.6938, 0.7043) \\
                   & $\hat{\sigma}$ & (0.0094, 0.0110) & (0.0902, 0.1048) & (0.2498, 0.2893)  \\
     0.8         & $\hat{\alpha}$  & (0.7999, 0.8004) & (0.7965, 0.8019) & (0.7946, 0.8082) \\
                   & $\hat{\sigma}$ & (0.0088, 0.0103) & (0.0926, 0.1071) & (0.2309, 0.2673)  \\
     0.9         & $\hat{\alpha}$  & (0.8997, 0.9005) & (0.8979, 0.9051) & (0.8953, 0.9146) \\
                   & $\hat{\sigma}$ & (0.0098, 0.0114) & (0.0939, 0.1088) & (0.2458, 0.2847)  \\ \hline
    \end{tabular}\label{tbl:robust}
  \end{center}
\end{table}

Additionally, as this paper is not particularly concerned about the extremes to the values of $\alpha$, the Beta distribution was used, and to further check robustness, $\alpha$* and $\beta$* were increased each to 5, 10, 20, 50, etc, to make sure the 95\% credible interval would still capture both $\alpha$* and $\beta$*. As seen in Table \ref{tbl:robustab} when this test was run, $\alpha$* and $\beta$* were both at 100 before the interval failed to capture the true value, in this case $\alpha$. To show robustness, both $\alpha$ and $\beta$ are changed, starting at $\alpha=1$ and $\beta=1$, also known as a uniform distribution, and increased at an equal rate until the 95\% credible interval no longer captured the parameters of the equation successfully. As is evident in the table above, the properties of the method derived capture the parameters in the case of a uniform distribution and continued to capture the parameters until the beta distribution was at $\alpha$ and $\beta$ both equal to 100.  Even under a very strict beta distribution, the properties of the method capture the parameters.

\begin{table}
  \begin{center}
   \caption{Robustness of estimating $\alpha = 0.82$ and $\sigma = 0.1$ study for simulated datasets with changing the parameters of the beta distribution, changing both the $\alpha$* and $\beta$* value. Quantiles of the posterior distribution were used to create 95\% credible intervals for $\alpha$ and $\sigma$, $(Q_{0.025},Q_{0.975})$. } 
    \begin{tabular}{ c | c | c  } \hline
     $\alpha$* \& $\beta$* &   $\alpha = 0.82$                       &  $\sigma = 0.1$  \\ \hline
     1         & $\hat{\alpha}$  & (0.8174, 0.8237) \\
                   & $\hat{\sigma}$ & (0.0984, 0.1143)  \\
     3         & $\hat{\alpha}$  & (0.8184, 0.8244) \\
                   & $\hat{\sigma}$ & (0.0982, 0.1141)  \\
     5         & $\hat{\alpha}$  & (0.8190, 0.8249)  \\
                   & $\hat{\sigma}$ & (0.0958, 0.1115) \\
     10         & $\hat{\alpha}$  & (0.8176, 0.8230)  \\
                   & $\hat{\sigma}$ & (0.0908, 0.1046)  \\
     20         & $\hat{\alpha}$  & (0.8165, 0.8220)  \\
                   & $\hat{\sigma}$ & (0.0853, 0.1000)  \\
     50         & $\hat{\alpha}$  & (0.8141, 0.8201) \\
                   & $\hat{\sigma}$ & (0.0920, 0.1065) \\
     100         & $\hat{\alpha}$  & (0.8202, 0.8260)   \\
                   & $\hat{\sigma}$ & (0.0894, 0.1042)   \\ \hline
    \end{tabular}\label{tbl:robustab}
  \end{center}
\end{table}

To further expound upon the robustness of the algorithm, the algorithm used 200 unique data sets to ensure that the algorithm successfully captured both $\alpha$ and $\beta$ at least $95$\% of the iterations tested. 

\begin{equation}
\hat{P}(C_\alpha) = \frac{\sum_{1}^{m}I_{\alpha_0\in C_\alpha}}{m}
\end{equation}
where $\alpha_0$ is an indicator variable, indicating whether or not the $\alpha$ was inside the $95\%$ credible interval, and the total amount of $\alpha$'$s$ in the interval divided by the total number of samples yielded a percentage, indicating coverage probability.

\begin{table}
  \begin{center}
   \caption{Coverage probabilities $\hat{P}(\alpha)$ for estimating $\alpha$ and $\sigma$ study for simulated datasets with $\alpha = 0.1, 0.3, 0.5, 0.7, and 0.9$ and $\sigma = 0.01, 0.1$ and $0.25$. Quantiles of the posterior distribution were used to create 95\% credible intervals for $\alpha$ and $\sigma$, $(Q_{0.025},Q_{0.975})$. Table indicates what percentage of $\hat{\alpha}$ contained based on 200 MCMC simulations. } 
    \begin{tabular}{ c c |  c c c  } \hline
     $\alpha$ &                         &  $\sigma = 0.01$ &  $\sigma = 0.1$  &  $\sigma = 0.25$ \\ \hline
     0.1         & $\hat{\alpha}$  & 0.995 & 0.980 & 0.965 \\
                   & $\hat{\sigma}$ & 0.995 & 0.980 & 0.975  \\
     0.3         & $\hat{\alpha}$  & 0.995 & 0.975 & 0.965 \\
                   & $\hat{\sigma}$ & 0.995 & 0.965 & 0.955  \\
     0.5         & $\hat{\alpha}$  & 1.000 & 0.980 & 0.965 \\
                   & $\hat{\sigma}$ & 0.995 & 0.975 & 0.965  \\
     0.7         & $\hat{\alpha}$  & 0.990 & 0.975 & 0.955 \\
                   & $\hat{\sigma}$ & 1.000 & 0.975 & 0.950  \\
     0.9         & $\hat{\alpha}$  & 0.995 & 0.985 & 0.960 \\
                   & $\hat{\sigma}$ & 0.995 & 0.990 & 0.955  \\ \hline
    \end{tabular}\label{tbl:robust}
  \end{center}
\end{table}
 
\newpage
\section{Conclusion}
Proof of concept for accurate estimation of the fractional parameter $\alpha$ with varying external error was concluded. This work demonstrates a method for researchers to employ fractional advection-diffusion equations on real world problems where the fraction can be estimated using observed data. This approach, while computationally intensive, not only can estimate the fraction but can also adequately quantify the uncertainty associated with the parameters and predictions. The method is shown to be robust with respect to the underlying value of the fraction. By testing the robustness of changing the true value of $\alpha$ and $\sigma$, the approach created $95\%$ credible intervals capturing the true value of $\alpha$ and $\sigma$. Furthermore, by changing the prior distribution to a Uniform distribution or Beta distribution with varying $\alpha$ and $\sigma$, the method is robust to capture the true values. By running the 200 MCMC simulations, the $95\%$ coverage probabilities successfully captures both parameters at least $95\%$ of the time. Utilizing the Sample Importance Resampling Regime and then tuning the importance sampler, the control of the quality of the sample ensured the differential equation model was solved correctly.

\section{Future Research}
This work has demonstrated that the Bayesian framework will allow for accurate estimation of $\alpha$ under the scenario of external error; a similar study should be performed under internal error.  Due to the stochastic nature of this scenario, the techniques for estimation may differ greatly such as using Metropolis-Hasting sampling possibly combined with Slice sampling.  This work has been omitted this paper as it is drastically different and the work performed loses focus.  The study should be a simulation study and verify that the methods can be combined to produce accurate results.  Ultimately, future research should focus on developing a Bayesian estimation approach when both the {\em internal and external error} exist in a fractional differential equation system. In addition to error, future research on scenarios with more parameters such as rock porosity, hydrocarbon density, hydrocarbon viscosity, and rock permeability to better model the transport of unconventional hydrocarbon reservoirs. Lastly, applying varying numerical and hybrid methods to test for parameter redundancy by adding constraints to result in better modeling in the future will impact the effectiveness of models. \\ \\

{\noindent{\bf Competing Interests}} \\ \\
The authors declare that there is no conflict of interests regarding the publication of this paper.

\end{document}